\documentclass[aps,prd,twocolumn,superscriptaddress,preprintnumbers,floatfix,nofootinbib,longbibliography,showlabels]{revtex4-1}
\pdfoutput=1

\usepackage[utf8x]{inputenc}
\usepackage{float}
\usepackage{graphicx}
\usepackage{color} 
\usepackage{hyperref}
\usepackage{epsfig}
\usepackage{wrapfig}
\usepackage{amsmath}
\usepackage{amssymb}
\newcommand{\tr}{\textrm{Tr}\,}

\newcommand{\msbar}{\overline{\mbox{\rm MS}}}

\def\lsim{\raise0.3ex\hbox{$<$\kern-0.75em\raise-1.1ex\hbox{$\sim$}}}
\def\gsim{\raise0.3ex\hbox{$>$\kern-0.75em\raise-1.1ex\hbox{$\sim$}}}

\newcommand{\beq}{\begin{eqnarray}}
\newcommand{\eeq}{\end{eqnarray}}
\newcommand{\be}{\begin{equation}}
\newcommand{\ee}{\end{equation}}

\newcommand{\bmat}{\left (\begin{array}{cc}}
\newcommand{\emat}{\end{array} \right )}

\definecolor{grey}{RGB}{100,100,100}

\newcommand{\half}{\mbox{${\frac12}$}}

\newcommand{\pd}{\partial}
\newcommand{\rmd}{\mathrm{d}}
\newcommand{\A}{\alpha}
\newcommand{\B}{\beta}
\newcommand{\C}{\gamma}
\newcommand{\D}{\delta}

\begin{document}

\title{Electroweak Sphaleron in a Magnetic field}

\author{Jaakko Annala}
\email{jaakko.annala@helsinki.fi}
\affiliation{Department of Physics and Helsinki Institute of Physics, PL 64 (Gustaf H\"allstr\"omin katu 2), FI-00014 University of Helsinki, Finland}
\author{Kari Rummukainen}
\email{kari.rummukainen@helsinki.fi}
\affiliation{Department of Physics and Helsinki Institute of Physics, PL 64 (Gustaf H\"allstr\"omin katu 2), FI-00014 University of Helsinki, Finland}

\date   {\today}

\begin{abstract}

Using lattice simulations we calculate the rate of baryon number violating processes, the sphaleron rate, in the Standard Model with an external (hyper)magnetic field for temperatures across the electroweak crossover, focusing on the broken phase. Additionally, we compute the Higgs expectation value and the pseudocritical temperature.
The electroweak crossover shifts to lower temperatures with increasing external magnetic field, bringing the onset of the suppression of the baryon number violation with it.  When the hypermagnetic field reaches the magitude $B_Y \approx 2 T^2$ the crossover temperature is reduced from $160$ to $145$\,GeV.
In the broken phase for small magnetic fields the rate behaves quadratically as a function of the magnetic flux. For stronger magnetic fields the rate reaches a linear regime which lasts until the field gets strong enough to restore the electroweak symmetry where the symmetric phase rate is reached.
\end {abstract}

\maketitle

\section{Introduction}

The results from the ATLAS and CMS experiments at the LHC are in complete agreement with the Standard Model of particle physics: a Higgs boson with a mass of $\approx 125$\,GeV has been discovered \cite{Aad:2012tfa,Chatrchyan:2012ufa}, and no evidence of beyond-the-Standard-Model physics has been observed.  If the electroweak-scale physics 
is fully described by the Standard Model, then the electroweak symmetry breaking transition in the early Universe was a smooth crossover from the symmetric phase at $T > T_c$, where the expectation value of the Higgs field was approximately zero, to the broken phase at $T < T_c$ where it is finite, reaching the value $246/\sqrt{2}$\,GeV at zero temperature.

The infrared problems inherent in high-temperature gauge theories \cite{Linde:1978px,Gross:1980br} make the physics nonperturbative. The overall nature of the transition was resolved already in the 1990s using lattice simulations \cite{Kajantie:1996mn,Gurtler:1997hr,Csikor:1998eu,Rummukainen:1998as}, which indicated that the transition is first order with Higgs masses $\lsim 72$\,GeV, and crossover otherwise. More recently, the precise thermodynamics of the crossover at the physical Higgs mass was analyzed in Ref.~\cite{DOnofrio:2015gop} (see also \cite{Laine:2015kra}), and e.g. the crossover temperature was determined to be $T_c = 159.6 \pm 1.5$\,GeV.

The chiral anomaly of the electroweak interactions lead to the non-conservation of the baryon and the lepton number \cite{tHooft:1976rip}.  In {\it electroweak baryogenesis} scenarios \cite{Kuzmin:1985mm,Rubakov:1996vz} the baryon number of the Universe arises through processes at a first order electroweak phase transition, and a smooth crossover makes these ineffective.  Thus, in electroweak baryogenesis the origin of the baryon asymmetry must be due to beyond-the-Standard-Model physics (for reviews, see e.g.~\cite{DiBari:2021fhs,Morrissey:2012db}).

The Chern-Simons (CS) number for weak SU(2) gauge is defined as
\begin{align}
  N^W_{\rm CS}(t)&\equiv \frac{g^2}{32\pi^2}\int_0^t \rmd t \int \rmd^3x \epsilon_{\A\B\C\D} \tr F^{\A\B} F^{\C\D} \ ,
\end{align}
where $F^{\A\B}$ is the field strength tensor of SU(2), $g$ is the SU(2) gauge coupling and $\epsilon_{\A\B\C\D}$ is the totally antisymmetric tensor.  Analogous to SU(2), the hypercharge U(1) CS number is given by
\begin{equation}
  N^Y_{\rm CS}(t) \equiv \frac{g^\prime{}^2}{32\pi^2}\int_0^t \rmd t \int \rmd^3x \epsilon_{\A\B\C\D}B^{\A\B}B^{\C\D}\ ,
\end{equation}
where $g'$ is the hypercharge gauge coupling.  The chiral anomaly couples the baryon and lepton numbers to the change in the Chern-Simons numbers as
\begin{equation}
  \Delta B = \Delta L = 3\Delta N_{\rm CS}\ ,
\end{equation}
where
\begin{align}\label{ncs_su2u1}
    &N_{\rm CS}(t) \equiv N^W_{\rm CS}(t)-N^Y_{\rm CS}(t)\ .
\end{align}
For SU(2), the Chern-Simons number is topological, and there exists infinitely many classically equivalent but topologically distinct vacua that cannot be continuously transformed into one another without crossing an energy barrier.  The {\it sphaleron} is a saddle point finite energy solution of the classical field equations separating two topologically distinct vacua \cite{Manton:1983nd, Klinkhamer:1984di}.  The CS number is an integer for vacuum field configurations and a half integer $N^W_{\rm CS} = \tfrac{1}{2} + n,\ n\in \mathbb{Z}$ for sphaleron configurations \cite{Klinkhamer:1984di}.

In contrast, the U(1) field has trivial topology and without an external (hyper)magnetic field its CS number in vacuum vanishes. However, in an external magnetic field 
the vacuum is degenerate with respect to the U(1) CS number which can obtain any value in contrast to the SU(2) case where it is an integer \cite{Giovannini:1997eg,Joyce:1997uy}.  This can lead to baryon and lepton number change on its own \cite{Giovannini:1997eg,Joyce:1997uy,Figueroa:2017hun,Figueroa:2019jsi,Kamada:2016eeb,Kamada:2018tcs}.  

Close to thermal equilibrium the evolution of the CS number is diffusive and is described by a diffusion constant known as the sphaleron rate
\begin{equation}\label{diff_rate}
  \Gamma = \lim_{V,t\to\infty}\frac{\langle  N_{\rm CS}(t)^2 \rangle}{V t} \ .
\end{equation}
In the absence of hypermagnetic fields the contribution of the U(1) can be neglected due to it having little effect on the form of the phase transition \cite{Kajantie:1996mn,DOnofrio:2015gop,Laine:1998jb} and the sphaleron rate \cite{Klinkhamer:1984di,Kleihaus:1991ks,Klinkhamer:1990fi,Kunz:1992uh}. 
In this framework the sphaleron rate has been studied extensively with analytical and numerical lattice methods. The general behavior is that in the broken phase the rate is suppressed by the energy of the sphaleron $\Gamma_{\text{brk}} \sim \alpha_W^4 T^4 e^{-E_{sph}/T}$ \cite{PhysRevD.36.581,Moore:1998swa} and in the symmetric phase the rate is unsuppressed behaving as $\Gamma_{\text{sym}} \sim \ln(1/\alpha_W) \alpha_W^5 T^4$ \cite{Arnold:1996dy,Bodeker:1998hm,Arnold:1998cy,Moore:1998zk,Moore:1999fs,Bodeker:1999gx}, where $\alpha_W = g^2/(4\pi)$.
In the Standard Model with the physical Higgs mass the sphaleron rate was recently measured using lattice simulations across the crossover from the symmetric phase to deep in the broken phase \cite{D'Onofrio:2014kta}.  The temperature where the transitions decouple, i.e. the baryon number freezes, was found to be $\approx 132$\,GeV, substantially below the crossover temperature $\approx 160$\,GeV.

The presence of a U(1) hypercharge magnetic field can affect both the thermodynamics of the crossover and the sphaleron rate. Large scale magnetic fields exist in the Universe which may have primordial origin, see e.g. reviews \cite{Subramanian:2015lua,Vachaspati:2020blt}. Primordial magnetic fields could have been generated before the electroweak transition corresponding to hypermagnetic fields before the transition which turn into the U(1)${}_{\text{em}}$ magnetic fields after the transition. The magnitude of such fields are largely unconstrained \cite{Durrer:2013pga}. However, see \cite{Kamada:2020bmb} for recent stronger constraints at larger scales.

When the U(1) is taken into account the spherical symmetry of the sphaleron reduces to axial symmetry and the sphaleron has a magnetic dipole moment. (It has been shown to be formed from magnetic monopole-antimonopole pair and a loop of electric current \cite{Hindmarsh:1993aw}.) Thus the minimum energy of the sphaleron can be lowered by an external magnetic field. In a small external field analytical estimates give a simple dipole interaction $\Delta E_{sph}=-\vec{B}_{ext} \cdot \vec{\mu}_{sph}$ \cite{Comelli:1999gt}. In addition the form of the phase transition is modified by an external magnetic field \cite{Kajantie:1998rz} which has an effect on the sphaleron rate through the transition.

At zero temperature the classical sphaleron energy has been computed on the lattice for a wide range of magnetic field values \cite{Ho:2020ltr}. The situation is complicated by the appearance of Ambjorn-Olesen phase for large magnetic field values. At a critical field value $B_{c1}=m_W^2/e$ the ground state becomes a nontrivial vortex structure and at a second critical value $B_{c2}=m_H^2/e$ the electroweak symmetry is restored \cite{Ambjorn:1988fx,Ambjorn:1988gb,Ambjorn:1989bd,Chernodub:2022ywg}. The sphaleron energy is found to decrease until at the second critical field value when the symmetry is restored the energy vanishes \cite{Ho:2020ltr}. At finite temperature around the electroweak scale previous studies have not been able to find the aforementioned vortex phase \cite{Kajantie:1998rz}.

Elaborate methods have been developed to compute the sphaleron rate on the lattice accurately \cite{Bodeker:1999ey,Moore:1998swa,Moore:2000mx,D'Onofrio:2012jk}. We employ the dimensionally reduced effective theory of the Standard Model and perform the first dynamical simulations of the sphaleron rate which includes the U(1) field and compute the sphaleron rate for different magnitudes for the external hypermagnetic field over the electroweak crossover with focusing on the behavior in the broken phase.

The structure of the paper is as follows. In Sec. \ref{sec:efftheory} we describe the effective theory and its lattice formulation that we will use in our simulations. In Sec. \ref{sec:measuring} the methods used to measure the sphaleron rate from the lattice is described. In Sec. \ref{sec:results} we present the results and finally in Sec. \ref{sec:conc} we conclude.

\section{Effective three-dimensional theory}\label{sec:efftheory}

In our simulations we use dimensionally reduced three-dimensional effective theory of the Standard Model. 
The method of dimensional reduction is made possible due to the fact that in finite temperature the fields are naturally expressed in terms of three-dimensional Matsubara modes having thermal masses around $\pi T$. This and the fact that Standard Model couplings are sufficiently small around the electroweak scale gives rise to a parametric hierarchy of scales in the Euclidean path integral $\pi T$, $gT$, $g^2 T$, called superheavy, heavy and light scales respectively. This allows us to integrate out the superheavy and heavy modes by well defined perturbative methods.  All the fermionic modes are integrated out since their Matsubara frequencies $\omega^f = (2n+1)\pi T$ are all proportional to $\pi T$. In addition all temporal bosonic modes $\omega^b = 2\pi n T$, $n\neq 0$ are also integrated out. Thus we are left with a 3d purely bosonic effective theory with the soft scales $g^2T$. The soft scales have to be studied regardless with nonperturbative methods due to the infrared problem in thermodynamics of Yang-Mills fields \cite{Linde:1980ts}. The resulting (super-)renormalizable Lagrangian reads
\begin{align}
  L =& \frac 1 4 \tr F_{ij} F_{ij} + \frac 1 4 B_{ij}B_{ij} \nonumber \\
  &+ (D_i\phi)^\dagger D_i\phi+m_3^2\phi^\dagger\phi + \lambda_3(\phi^\dagger\phi)^2, \label{continuum_theory}
\end{align}
where
\begin{align}
  F_{ij} &= \partial_iA_j -\partial_jA_i - g_3
  [A_i, A_j],  ~~~
  A_i =  \half\sigma_aA_i^a \nonumber\\
  B_{ij} &= \partial_iB_j-\partial_jB_i    \\
  D_i&= \partial_i+ig_3A_i+ig'_3B_i/2.  \nonumber
\end{align}
Here $A_i$,$B_i$ are the 3d SU(2) and U(1) gauge fields; $g_3$, $g_3^\prime$ are the dimensionful SU(2) and U(1) couplings and $\phi$ is a complex scalar doublet. 

The dimensionful parameters of the 3d theory $g_3, g_3^\prime, \lambda_3, m_3^2$ are mapped to Standard Model parameters $\alpha_S, G_F, m_H, m_W, m_Z, m_{t}$ and the temperature $T$ via a perturbatively computable functions. All the details of the construction of the effective 3d theory and the mapping of parameters can be found in Refs. \cite{Farakos:1994xh, Farakos:1994kx, generic}. The accuracy of the 3d effective theory has been estimated to be $ \sim 1 \% $ \cite{Jakovac:1994xg,Farakos:1994kx,Farakos:1994xh,generic,Laine:1999rv}.

\begin{figure}[t]
  \includegraphics[width=1.0\columnwidth]{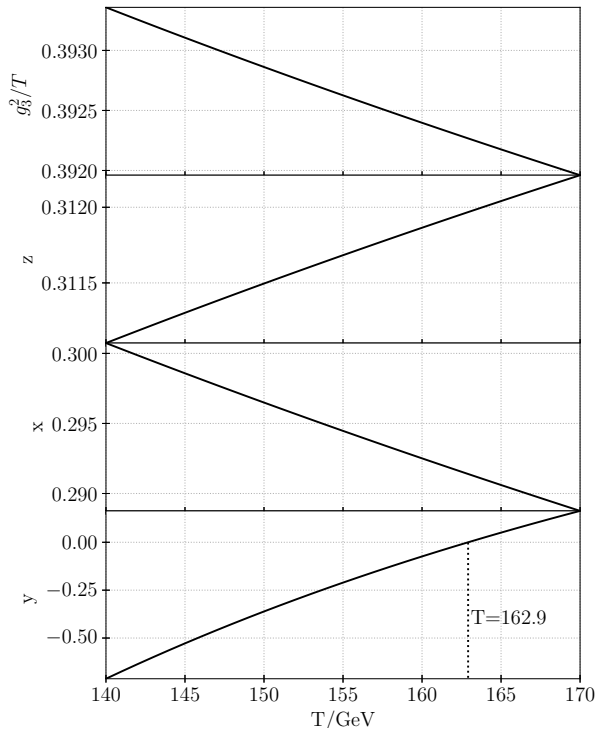}
  \caption{The temperature dependence of the dimensionless effective 3d theory parameters using the Standard model parameters $G_F=1.1663788\times 10^{-5}\,\text{GeV}^{-2}$, $m_H = 125.25$\,GeV, $m_Z = 91.1876$\,GeV, $m_W = 80.377$\,GeV, $m_{t} = 172.69$\,GeV, $\alpha_S = 0.1179$ \cite{Workman:2022ynf}.}
  \label{fig:xyz}
  \vspace{-0mm}
\end{figure}

We choose the SU(2) coupling $g_3^2$ to set the scale and use a set of dimensionless couplings defined by
\begin{equation}
  x\equiv \frac{\lambda_3}{g_3^2},\qquad
  y\equiv \frac{m_3^2}{g_3^4},\qquad
  z\equiv \frac{g_3'^2}{g_3^2}.
  \label{3d_params}
\end{equation}
The three parameters and the scale are plotted in Fig.~\ref{fig:xyz} in the relevant temperature range (a code for computing these parameters can be found in zenodo \cite{zenodo}). As seen from the plot only the parameter $y$ varies significantly over temperature and it is the natural choice for the temperature variable of the system. In \cite{DOnofrio:2015gop} the crossover temperature (defined as the peak of the susceptibility of the Higgs condensate) was found to be few GeV below the temperature where $y=0$. We find $y=0$ at $T = 162.9$\,GeV which is slightly different from \cite{DOnofrio:2015gop} due to using an updated value for the top mass.

\subsection{Lattice action}\label{sec:lattice}

The 3d effective theory in purely bosonic and straightforward to put on the lattice. For convenience we write the Higgs field as
\begin{equation}
\Phi= \frac{1}{g_3^2} \biggl((\tilde\phi)(\phi)\biggr)\equiv
\frac{1}{g_3^2} 
\left(\begin{array}{cc}
    \phi_2^*&\phi_1\\
    -\phi_1^*&\phi_2
  \end{array}\right),
\label{phimatrix2}
\end{equation}
which transforms under the SU(2)$\times$U(1) gauge transformation as 
\begin{equation}
\Phi(x)\to G(x)\Phi e^{-i\theta(x)\sigma_3},
\label{gt}
\end{equation}
where $\sigma_3$ is the third Pauli matrix and $G(x)$ is an element of SU(2). Now the lattice action that corresponds to the continuum theory \eqref{continuum_theory} can be written as
\begin{align}\label{lattice_action}
S&= \beta_G \sum_x \sum_{i<j}[1-\half \tr P_{ij}] 
  +  \beta_Y \sum_x \sum_{i<j} \half \alpha_{ij}^2 \nonumber\\
 &- \beta_H \sum_x \sum_{i}
\half\tr\Phi^\dagger(x)U_i(x)\Phi(x+i)
e^{-i\alpha_i(x)\sigma_3}     \\
 &+ \beta_2 \sum_x
 \half \tr\Phi^\dagger(x)\Phi(x) + \beta_4\sum_x
 \big[\half\tr\Phi^\dagger(x)\Phi(x)\big]^2, \nonumber
\end{align}
where
\begin{align}
  P_{ij}(x) &= U_i(x)U_j(x+\hat i)
  U^\dagger_i(x+\hat j)U^\dagger_j(x), \label{su2plaq} \\
  \alpha_{ij}(x) &=\alpha_i(x)+\alpha_j(x+\hat i)-
  \alpha_i(x+\hat j)-\alpha_j(x). \label{u1plaq}
\end{align}
Here $U_i(x), \alpha_i(x)$ are the SU(2) and noncompact U(1) link variables, respectively, and $P_{ij}, \alpha_{ij}$ are their corresponding plaquettes. The lattice parameters $\beta_G, \beta_Y, \beta_H, \beta_2, \beta_4$ are related to the continuum parameters by perturbatively computable functions computed in \cite{Laine:1997dy}. In addition we employ the partial $O(a)$ improvements on these relations \cite{Moore:1996bf, Moore:1997np}. Notably, $\beta_G = 4/(g_3^2 a) + 0.6674...$ with $a$ being the lattice spacing. Rest of the lengthy relations can be found in Appendix \ref{appendix_cont_rel}.

Now the lattice observable $\langle \half\tr\Phi^\dagger\Phi \rangle$ is related to the $\msbar$ renormalized 3d continuum value $\langle\phi^\dagger\phi\rangle$ by \cite{Laine:1997dy}
\begin{align}
  \frac{\langle\phi^\dagger\phi\rangle}{g_3^2}= &Z_g Z_m\Big[
  \langle \half\tr\Phi^\dagger\Phi \rangle
  -\frac{\Sigma\beta_G}{8\pi} \nonumber \\
  &-\frac{3+\bar{z}}{16\pi^2}\Bigl(\log( 3\beta_G/2)
  +0.6679... \Bigr)\Big],\label{rl2}
\end{align}
where $Z_g,Z_m$ and $\bar{z}$ are defined in Appendix \ref{appendix_cont_rel}.
Finally, the 3d expectation value is related to the physical SM Higgs expectation value $v$ as 
\begin{equation}
  v^2/T^2 = 2\langle \phi^\dagger \phi \rangle/T .
\end{equation}

\subsection{Hypermagnetic field on the lattice}

A flux of magnetic field perpendicular to the $x_3$ axis,
\begin{equation}
  g_3^\prime \Phi_B = \int \rmd x_1 \rmd x_2 B_{12}(x),
\end{equation}
can be imposed to the lattice by modifying the periodic boundary conditions of the U(1) link variables $\alpha_i$ \cite{Kajantie:1998rz}. Requiring the action to be periodic quantizes the total flux $g_3^\prime \Phi_B/2 = 2\pi n_b$, $n_b\in \mathbb{N}$. Without this restriction there would be boundary defects and the translational invariance would be lost. One possible way to add a flux of magnitude $g_3^\prime \Phi_B/2 = 2\pi n_b$ is by modifying the boundary conditions as 
\begin{equation}
  \alpha_1(n_1,0,n_3) - \alpha_1(n_1,L_2,n_3) = 2\pi n_b \delta_{n_1,1} \ ,
\end{equation}
for each $n_3$ in a lattice with extent $L_1 L_2 L_3$. We define a dimensionless parameter describing the average magnetic flux density as
\begin{equation}\label{flux_b}
  b \equiv \frac{g_3^\prime B_Y^{3d}}{g_3^4} = \frac{4\pi n_b}{L_1 L_2}\left(\frac{1}{g_3^2 a}\right)^2 \ ,
\end{equation}
where $B_Y^{3d} \equiv \Phi_B / (L_1 L_2)$ is the magnetic flux density which is related to the four-dimensional density as $B^{3d}_{Y} \simeq B_Y^{4d}/\sqrt{T} + \mathcal{O}(g^\prime{}^2)$. The dimensionless parameter then relates to the $4d$ flux approximately as $B_Y^{4d} = (g^\prime/g^4) b T^2 + \mathcal{O}(g^\prime{}^3)$. We cannot use the effective 3d theory to simulate arbitrarily large magnetic fields due to the external magnetic field affecting higher dimensional operators invalidating the effective theory, so we require $b \ll 2 \pi^2$ \cite{Kajantie:1998rz}.

\section{Measuring the Sphaleron rate}\label{sec:measuring}

\subsection{Real time evolution}\label{sec:evolution}
In itself the effective 3d theory \eqref{continuum_theory} does not describe dynamical phenomena, such as the sphaleron process. As shown by Arnold, Son and Yaffe \cite{Arnold:1996dy}, the classical equations of motion suffer from ultraviolet divergences which prevent taking the continuum limit on the lattice.

However, in SU(2) gauge theory the dynamics of the soft modes ($k \lesssim g^2 T$), which are relevant for sphaleron transitions, are fully overdamped and to leading logarithmic accuracy $1/\ln(1/g)$ the evolution can be described by Langevin equation with Gaussian noise $\xi^a_i$ (in $A_0 = 0$ gauge) \cite{Bodeker:1998hm,Bodeker:1999ey,Arnold:1999jf,Arnold:1998cy}:
\begin{align}\label{langevin}
  \pd_t A_i &= -\frac{1}{\sigma_{el}}\frac{\pd H}{\pd A_i} + \xi^a_i \ , \\
  \langle \xi^a_i(x,t)\xi^b_j(y,t^\prime)\rangle &= 2\sigma_{el}T \delta_{ij}\delta^{ab} \delta^3(x-y)\delta(t-t^\prime) \ ,
\end{align}
where $H/T = S$ with $S$ defined in \eqref{lattice_action}. Here $\sigma_{el} \simeq 0.9239  T$ \cite{Arnold:1999uy} is the non-Abelian color conductivity of SU(2). 

It can be shown that any diffusive field update algorithm, for example the heat bath update, is equivalent to Langevin evolution \cite{Moore:1998zk}. This is advantageous because heat bath update is computationally much more efficient. The Langevin time $t$ for SU(2) can be related to performing $n$ full random order heat bath update sweeps as $\Delta t = \tfrac{1}{4}\sigma_{el} a^2 n$ and the leading corrections are observed to be small \cite{Moore:1998zk,Moore:2000jw}.
The heat bath approach enables us to take a well-defined continuum limit on the lattice.

The Higgs field evolves parametrically much faster than the SU(2) gauge field \cite{Moore:2000jw}.  Thus, the Higgs field almost equilibrates in the background of the instantaneous SU(2) field. This can be achieved by updating the Higgs field much more often than the SU(2) field.  We use a mixture of overrelaxation and heat bath updates, see \cite{nonpert} for details of the algorithms used. We increased the number of Higgs updates until the lattice observables of interest stayed constant resulting to around 50 more Higgs updates per gauge field update (similarly as in \cite{Gould:2022ran}).

\begin{figure}[ht]
  \includegraphics[width=1.0\columnwidth]{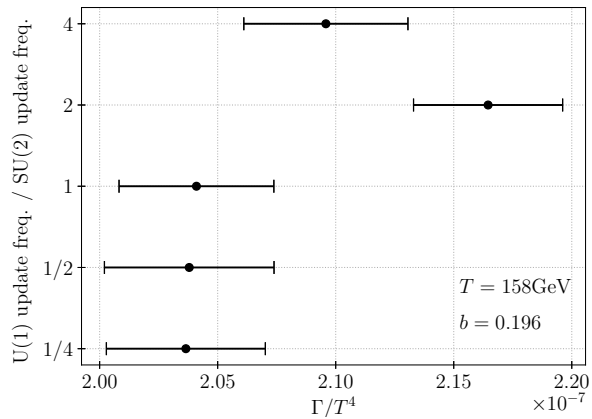}
  \caption{Comparing the obtained sphaleron rate with different U(1) update frequencies. On the y axis the ratio of U(1) updates per SU(2) update. The dependence is observed to be negligible with our statistical accuracy. }
  \label{fig:nu1freq}
  \vspace{-0mm}
\end{figure}

Finally, in the broken phase the U(1) field also evolves faster than the SU(2) gauge field for wavelengths relevant for sphaleron transitions.  The size of the sphalerons $\sim (g^2 T)^{-1}$ is given by the SU(2) dynamics.  Because the U(1) gauge coupling $g'^2$ is much smaller than the SU(2) coupling $g^2$, the U(1) modes with wavelength $\lambda \sim (g^2 T)^{-1}$ behave as weakly coupled nondamped modes evolving with timescale $\tau \approx \lambda$.  This is in contrast to the overdamped SU(2) evolution with timescale $\propto (\lambda^2)$.  Thus, on the lattice the sphaleron rate should be independent of the U(1) update rate provided it is frequent enough in comparison with the SU(2) updates. Indeed, we have tested this behavior with a few simulations with different heat bath update frequencies for the U(1) field and found no significant effect to the sphaleron rate, as seen from Fig.~\ref{fig:nu1freq}. In our final analysis we use equal update frequency for SU(2) and U(1) fields.

We note that in the broken phase the magnetic field remains unscreened and very long-range magnetic fields evolve very slowly in comparison with other fields (magnetohydrodynamics).  These modes have very small effect on the sphalerons, and indeed if there is no external (hyper)magnetic field the contribution from the U(1) sector is usually ignored \cite{D'Onofrio:2012jk,D'Onofrio:2014kta,Moore:1998swa}.

In the symmetric phase the SU(2) and U(1) Chern-Simons numbers are effectively decoupled and evolve independently.  The U(1) Chern-Simons number is not topological, and there is no characteristic length scale for its evolution when the external magnetic field is present.  It is not clear how to accurately capture the full quantum dynamics in numerical lattice simulations in this case.  However, this is not a problem for the analysis of the sphaleron rate in the broken phase, and, as will be discussed in Sec. \ref{sec:results}, the effect of the U(1) remains subleading in comparison with the SU(2) rate in the symmetric phase.

\subsection{Calibrated cooling}

Topology is not well defined on a discrete lattice, and a naive discretization of the CS number leads to ultraviolet noise which ruins the sphaleron rate measurement.  However, for sufficiently fine lattice spacing the sphaleron is large in lattice units with a length scale of order $1/(g^2T)$ \cite{Moore:1999fs}. This makes it possible to use methods which filter out the ultraviolet noise and allows us to accurately integrate the CS number.  One of these methods is the calibrated cooling \cite{Ambjorn:1997jz,Moore:1998swa}, which we employ here with the modification that we use {\it gradient flow} for all fields and integrate both the SU(2) and the U(1) CS numbers. Crucially, in the broken phase we track the difference of the CS numbers \eqref{ncs_su2u1}.  Periodically we cool all the way to the vacuum and check that the vacuum-to-vacuum integration result is close to an integer and remove any residuals in order to avoid the accumulation of errors.

Parametrizing the SU(2) links as $U_i(x)=\exp[i\theta^a_i(x)\sigma^a/2]$ the gradient flow can be written as
\begin{align}\label{grad_flow}
  \frac{\pd U_i(x)}{\pd \tau} &= - i \frac{\sigma^a}{2} U_i(x) \frac{\pd S}{\pd \theta^a(x)} \ ,\\
  \frac{\pd \alpha_i(x)}{\pd \tau} &= - \frac{\pd S}{\pd \alpha_i(x)} \ , \\
  \frac{\pd \Phi(x)}{\pd \tau} &= - \frac{\pd S}{\pd \Phi(x)} \ ,
\end{align}
where $\tau$ is the flow time. Evolving the fields with the gradient flow equations removes ultraviolet fluctuations smoothing the fields with a smoothing radius related to the flow time by $r=\sqrt{6\tau}a$ in three dimensions \cite{Luscher:2010iy}.

\begin{figure}[t]
  \includegraphics[width=1.0\columnwidth]{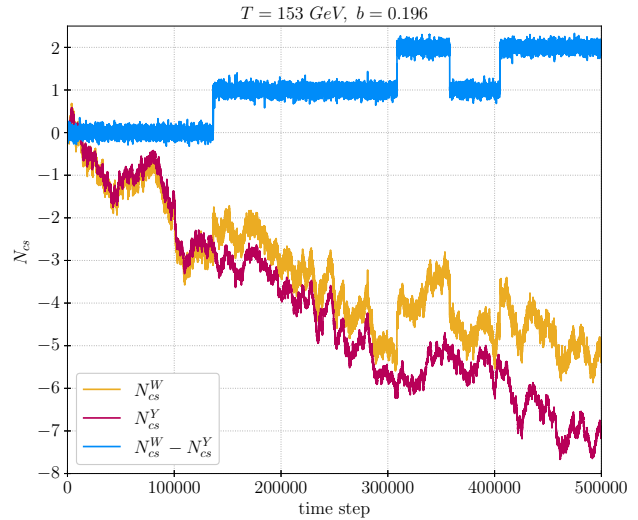}
  \caption{Real time CS trajectory in the broken phase at $T=153$\,GeV in an external magnetic field $b=0.196$.  SU(2) $N_{\rm CS}$ trajectory (yellow), U(1) $N_{\rm CS}$ trajectory (red) and their difference (blue).  It is clear that the difference becomes frozen at low temperatures.}
  \label{fig:ncs_walk}
  \vspace{-0mm}
\end{figure}

With these methods we can integrate the CS number accurately from a real time trajectory generated by heat bath updates. In the symmetric phase the SU(2) CS number diffuses rapidly between vacua. In the broken phase the SU(2) and U(1) gauge fields mix and the diffusion of the difference of the Chern-Simons numbers, $N_{\rm CS} = N_{\rm CS}^W - N_{\rm CS}^Y$, slows down dramatically, jumping between integer values.  This can be seen in Fig.~\ref{fig:ncs_walk}, where the CS number is measured somewhat below the crossover temperature.  Interestingly, the SU(2) and U(1) Chern-Simons numbers are not suppressed individually, only their difference is.  This is precisely the quantity which couples to the baryon and lepton number.

Finally, from the real time trajectory we can compute the sphaleron rate. We use the cosine transform method described in \cite{Moore:1997cr}.

\subsection{Multicanonical method}

Near the crossover temperature we measure the sphaleron rate using the real-time simulation methods discussed above, but
deep in the broken phase the rate gets strongly suppressed and normal methods become impractical. At any reasonable amount of simulation time only few transitions take place, if any. Thus in the broken phase we have to use special multicanonical methods to compute the rate. Details of the method can be found in \cite{Moore:1998swa,Moore:2000jw,D'Onofrio:2012jk}. The computation consists of two parts. The multicanonical method is used to measure the probabilistic suppression of the sphaleron at the height of the potential barrier, i.e. the $N_{\rm CS}$ distribution between two integer vacua $P(N_{\rm CS})$. In a nonzero magnetic field we need to use the $N_{\rm CS}$ given by \eqref{ncs_su2u1} so that in a vacuum it is an integer. Dynamical simulations are performed to compute the rate of tunneling over the top of the barrier. The tunneling rate is computed by measuring $|\Delta N_{\rm CS}/\Delta t|$ from dynamical simulations when the trajectory crosses the sphaleron barrier $N_{\rm CS}=\tfrac{1}{2}$. This needs to be compensated by a dynamical prefactor $\rmd = \sum_{\text{traj}} \delta_{\text{tunnel}}/(N_{\text{cross}}N_{\text{traj}})$ where $\delta_{\text{tunnel}}=0$ if the trajectory does not get to a new vacuum and $\delta_{\text{tunnel}}=1$ if it does, and $N_{\text{cross}}$ is the number of times $N_{\rm CS}$ crosses the barrier. This is needed due to the fact that the dissipative update is noisy which can result in multiple crossing of the barrier in a one trajectory. With these ingredients the sphaleron rate is given by
\begin{equation}
  \Gamma = \frac{P(|N_{\rm CS}-\tfrac{1}{2}|<\tfrac{\epsilon}{2})}{\epsilon V} \left\langle \left| \frac{\Delta N_{\rm CS}}{\Delta t} \right| \right\rangle \rmd \ ,
\end{equation}
where $\epsilon \ll 1$ (we used $\epsilon = 0.04$).

\section{Results}\label{sec:results}

We investigated the lattice spacing dependence of the sphaleron rate with an external hypermagnetic field for a few temperatures in the symmetric and broken phase. The parameters were chosen such that the external magnetic field had the same value, see Table \ref{tab:cl}.
\begin{table}[ht] 
  \begin{tabular}{ |c|c|c|c| } 
   \hline
   $4/(a g_3^2)$ & $~~~V/a^3~~~$ & $~~~n_b~~~$ &$~~~b~~~$ \\
   \hline
   5.6       & $16^3$  & 2     & 0.196 \\ 
   8         & $16^3$  & 1     & 0.196 \\ 
   10        & $20^3$  & 1     & 0.196 \\ 
   12        & $24^3$  & 1     & 0.196 \\ 
   \hline
  \end{tabular}
  \caption{Lattice spacings, volumes, magnetic flux and magnitude of the external magnetic field used when investigating the lattice spacing dependence. }
  \label{tab:cl}
\end{table}

\begin{figure}[t]
  \includegraphics[width=1.0\columnwidth]{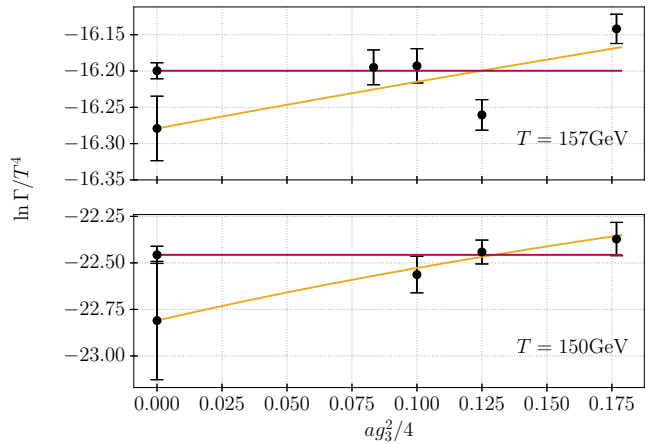}
  \caption{Sphaleron rate with few different lattice spacings $4/(a g_3^2)=5.6, 8, 10, 12$ in an external magnetic field $b=0.196$, see Table \ref{tab:cl}. Top most plot at $T=157$\,GeV close to the crossover using a normal simulation. Bottom plot at $T=150$\,GeV deep in the broken phase with multicanonical simulation. Both showing a linear and constant fits.}
  \label{fig:cl}
  \vspace{-0mm}
\end{figure}

In the symmetric phase and close to the crossover in the broken phase the lattice spacing dependence on the sphaleron rate is small, see the top most plot in Fig.~\ref{fig:cl}. This is similar to what was observed in previous studies without U(1) \cite{D'Onofrio:2012jk}.

In the broken phase for large lattice size and small lattice spacing even the multicanonical method becomes very inefficient, making the measurement of the Chern-Simons number evolution impractical at large lattices.
This prevents us from obtaining sufficient range in lattice spacings for a reliable continuum limit deep in the broken phase. Nevertheless, our limited results show only a mild lattice spacing dependence, as shown in Fig.~\ref{fig:cl}.  In the following most of our results have been obtained at single lattice spacing $g_3^2 a = 1/2$.

Similar inefficiency was noted in previous works where the U(1) field was omitted \cite{D'Onofrio:2012jk}. In our case the problem appears to be worse, presumably due to the additional noise of the combined SU(2) and U(1) Chern-Simons number observable.

We investigated the finite volume effects on the sphaleron rate in an external magnetic field with $b=0.196$ for a few different volumes $L^3a^3$ with $L = 8 / g_3^2, 13.9 / g_3^2, 16 / g_3^2$. The chosen temperatures were in the symmetric and in the broken phase near the crossover so that we could still use nonmulticanonical simulations. Similar to the previous studies we do not observe systematic volume dependence above $L = 8 / g_3^2$. In pure SU(2) theory it was found that $L=8/g_3^2$ is close to the smallest volume where the finite size effects are negligible \cite{Moore:1999fs}.

Due to the small observed lattice spacing dependence and no significant finite size effects at $L=8/g_3^2$, we present the results for the lattice parameters $g_3^2 a = 1/2$, $V=16^3 a^3$ when deep in the broken phase where we need to use the multicanonical simulations. With these parameters the lattice is still small enough for us to get reliable measurements of the CS number. This enables us to get good statistics with reasonable computational effort. Due to the magnetic field flux being quantized as \eqref{flux_b} the flux quanta are quite large for small volumes and we can only obtain a few different values of the magnetic flux for the multicanonical simulations. Thus in addition we present results for the lattice parameters $g_3^2 a = 1/2$, $V=32^3 a^3$ using nonmulticanonical simulations as deep as possible in to the broken phase.

For all nonmulticanonical runs we simulated $2 \times 10^6$ time steps; and for all multicanonical simulations we generated $12\times 10^3$ trajectories and generated $\sim 3 \times 10^6$ realizations to estimate the CS number distribution $P(N_{CS})$.

\subsection{Zero magnetic field}
\begin{figure}[t]
  \includegraphics[width=1.0\columnwidth]{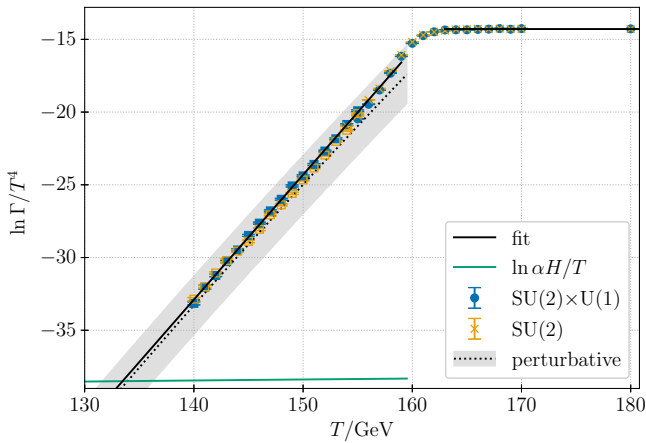}
  \caption{Sphaleron rate in the absence of the external magnetic field in theories with and without the U(1) field. Data points marked with squares are obtained using the multicanonical method. Fits are performed using the data that include U(1). The perturbative line is from \cite{Burnier:2005hp} with their nonperturbative correction removed.}
  \label{fig:rate_b0}
  \vspace{-0mm}
\end{figure}

Let us first present the results for zero external magnetic field since we find slightly different results as in previous works. We measure the sphaleron rate from simulations with and without the dynamical hypercharge U(1) field.  We do not observe any systematic difference between the results, see Fig.~\ref{fig:rate_b0}.  This justifies the omission of the U(1) field when there is no external magnetic field, as done e.g. in \cite{D'Onofrio:2014kta}.  Below we discuss results with the U(1) field included.

The Higgs field expectation value is observed to be very close to the perturbative result \cite{nonpert,Laine:2015kra} even without taking a continuum limit, see $b=0$ points in Fig.~\ref{fig:phi2}. For the Higgs expectation value it is straightforward to check the continuum limit because the Chern-Simons number measurement can be omitted. We measured the Higgs expectation value on lattice spacings $g_3^2 a = 1/2$, $1/3$ and $1/4$ on a few temperature values and found the continuum limit to match the perturbative result.

In the symmetric phase the measured sphaleron rate is approximately constant, with the value
\begin{equation}\label{sym_fit}
  \Gamma_{\text{sym.}}/T^4 = (6.23 \pm 0.05)\times 10^{-7} \approx (13.9 \pm 0.1)\alpha_W^5 \ ,
\end{equation}
with $\alpha_W\approx 0.03389$ at the electroweak scale.\footnote{The numerical factor in front of $\alpha_W^5$ includes contributions from logarithmic factors $\ln\alpha_W$ \cite{Bodeker:1998hm}. This form is presented for easier comparisons with earlier work.} In the broken phase the rate is well fitted by a pure exponential and we obtain
\begin{equation}\label{brk_fit}
  \ln(\Gamma_{\text{brk.}}/T^4) = (0.86 \pm 0.01)T/\text{GeV}  - (153.1 \pm 0.9 )\ . 
\end{equation}
Using the linear fit we can estimate when the sphaleron processes freeze out. This happens when the Hubble rate $H(T)$ becomes comparable to the sphaleron rate $\Gamma(T_*)/T_*^3 = \alpha(v/T_*)H(T_*)$. The function $\alpha(v/T_*)$ (where $v$ is the Higgs expectation value) is well approximated by a constant $\alpha = 0.1015$ in the relevant temperature range. Furthermore, $H(T)^2 = g_* \pi^2 T^2 /(90 M_{\text{Pl}}^2)$ where the effective number of degrees of freedom is well approximated by $g_*=106.75$ over the electroweak scale. The Hubble rate is seen in Fig.~\ref{fig:rate_b0} as the green line. With these we find the freeze-out temperature $T_* = 133.5 \pm 0.97$\,GeV.

The freeze-out temperature is slightly higher than the value obtained in Ref.~\cite{D'Onofrio:2014kta}. The differences are due to using an updated value for the top mass and the fact that the previous simulations did not fully implement the partial $O(a)$ improvement of the lattice parameters (see Appendix \ref{appendix_cont_rel}).
Because neither of the computations have been able to obtain a reliable continuum limit, the lack of improvement has an effect on the final results.\footnote{We note that the analysis of the thermodynamics of the Standard Model crossover in Ref.~\cite{DOnofrio:2015gop} implements the continuum limit, but the sphaleron rate is not measured.}

The main effect of both the improvement and the updated top mass is to effectively reduce the parameter $y$. The new top mass changes $y$ by an approximately constant shift $0.02$, whereas the partial $O(a)$ improvement modifies $y$ in a temperature-dependent manner, so that close to the pseudocritical temperature at $T \approx 160$\,GeV the effect is small but at $T \sim 140$\,GeV it reduces $y$ by $\sim 0.09$.  The net effect is that the sphaleron rate at $b=0$, Fig.~\ref{fig:rate_b0}, reaches a given value at slightly higher temperatures than in Ref.~\cite{D'Onofrio:2014kta}: for example, the almost-symmetric phase value $\ln \Gamma/T^4 = -16$ is reached at $T =158.8$\,GeV in \cite{D'Onofrio:2014kta} and we obtain $T=159.4$\,GeV here.  Deep in the broken phase the shift is slightly larger, value $\ln\Gamma/T^4 = -30$ is obtained at $T=141.9$\,GeV and $T=143.1$\,GeV, in \cite{D'Onofrio:2014kta} and here respectively. This difference is well within estimated systematic errors.

\subsection{Nonzero magnetic field}

Let us now look at the results for nonzero external magnetic field. We ran simulations with $g_3^2 a=1/2$ and volume $V=16^3a^3$ with magnetic flux quantum $n_b=0,1,2,3,4$ yielding $b$ in range from $b=0$ to $0.785$ with a step of $\Delta b=0.196$. To get a smaller step size for $b$ we additionally performed simulations with $g_3^2 a=1/2$ and volume $V=32^3a^3$ (without multicanonical simulations due to problems discussed above) with magnetic flux quantum $n_b=0,1,2,3,4$ and $6,8,10,12,14,16,18,20,22,24$ yielding a range $b=0$ to $1.178$ with $\Delta b = 0.049$.

\begin{figure}[t]
  \includegraphics[width=1.0\columnwidth]{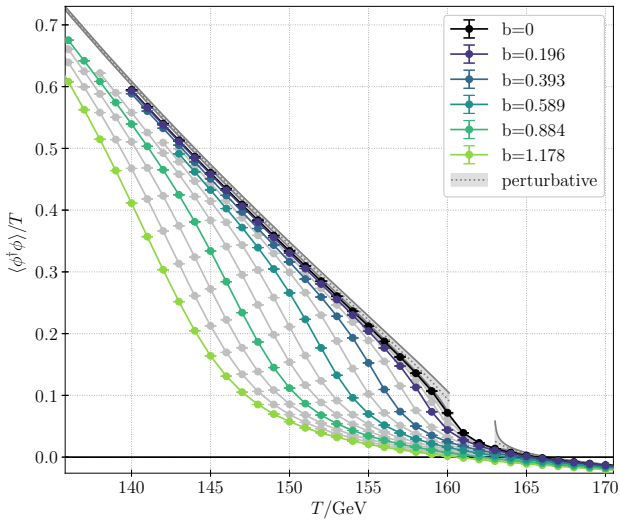}
  \caption{Higgs expectation value with different values for the magnetic field, with $V=32^3a^3$. The lines are added for clarity, they are not fits. The expectation value becomes negative in the symmetric phase due to additive renormalization factors, see \eqref{rl2}. The gray contours are the zero magnetic field symmetric \cite{Laine:2015kra} and broken phase \cite{nonpert} perturbative results.}
  \label{fig:phi2}
  \vspace{-0mm}
\end{figure}

\begin{figure}[t]
  \includegraphics[width=1.0\columnwidth]{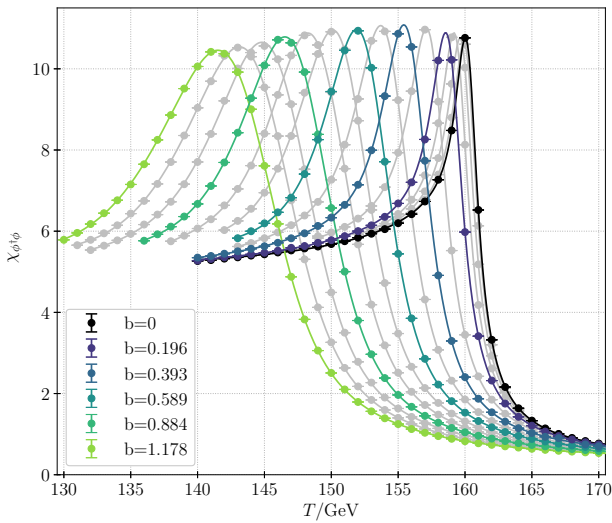}
  \caption{The dimensionless susceptibility with different magnitudes for the magnetic field, with $V=32^3a^3$. The lines are from fitting the interpolating function (defined in \cite{DOnofrio:2015gop}) to the data.}
  \label{fig:chi2}
  \vspace{-0mm}
\end{figure}

\begin{figure}[t]
  \includegraphics[width=1.0\columnwidth]{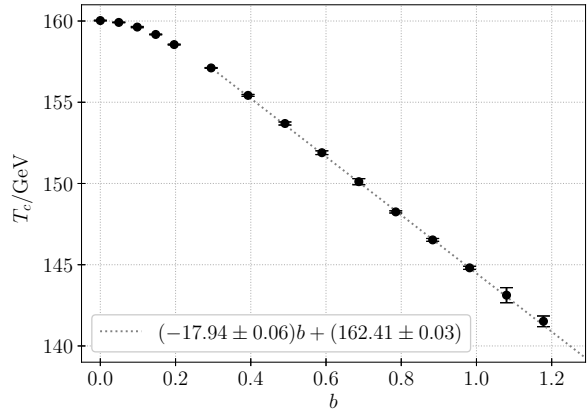}
  \caption{Pseudocritical temperature against the magnitude of the magnetic field.}
  \label{fig:Tcb}
  \vspace{-0mm}
\end{figure}

The form of the electroweak crossover is changed by the external magnetic field. This can be clearly seen from plotting the Higgs expectation value against the temperature with different magnitudes for the magnetic field, see Fig.~\ref{fig:phi2}. The crossover temperature can be seen to shift to smaller temperatures.
To get a better picture on the effect of the magnetic field on the crossover let us look at the susceptibility of the Higgs field. We define the crossover or pseudocritical temperature $T_c$ as the location of the maximum in the dimensionless susceptibility
\begin{equation}
  \chi_{\phi^\dagger\phi}(T) = VT\left\langle \left[ (\phi^\dagger\phi)_V - \left\langle \phi^\dagger\phi \right\rangle \right]^2  \right\rangle 
\end{equation}
where $(\phi^\dagger\phi)_V = 1/V \int_V \phi^\dagger \phi$ is the volume average. We use the interpolating function defined in \cite{DOnofrio:2015gop} to estimate the location of the peak. The susceptibility with different magnitudes of the magnetic field are shown in Fig.~\ref{fig:chi2}. From this we clearly see that the pseudocritical temperature is shifted to smaller temperatures and the crossover region gets wider in the sense of widening the peak of the susceptibility. The pseudocritical temperature against the magnitude of the magnetic field can be seen in Fig.~\ref{fig:Tcb}. With small field magnitudes it behaves quadratically after which it quickly reaches linear regime.  At $b=1$ ($B_Y^{4d} \approx 2T^2$)
the crossover temperature has decreased from $160$ down to $145$\,GeV.


Let us finally look at how the sphaleron rate is affected by a non-zero external magnetic field. 

\begin{figure}[t]
  \includegraphics[width=1.0\columnwidth]{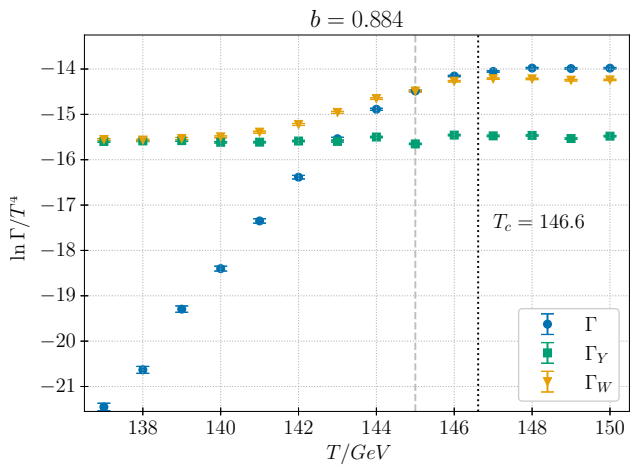}
  \caption{Example of diffusion rates of the pure SU(2) CS number, pure U(1) CS number and their difference \eqref{ncs_su2u1} for magnetic field magnitude $b=0.884$. Black dotted vertical line is the pseudocritical temperature.}
  \label{fig:su2_vs_su2u1}
  \vspace{-0mm}
\end{figure}

We measure the SU(2) and U(1) diffusion rates separately, and in Fig.~\ref{fig:su2_vs_su2u1} we show an example of the behavior of the rates through the crossover at $b= 0.884$. The U(1) diffusion rate $\Gamma_Y/T^4$ is seen to stay constant through the transition. In the high-temperature symmetric phase $N_{\rm CS}^W$ and $N_{\rm CS}^Y$ evolve independently, and the evolution of $N_{\rm CS}^W - N_{\rm CS}^Y$ (with rate $\Gamma/T^4$) is slightly faster than the evolution of each of the components alone. In the broken phase the Chern-Simons numbers are strongly correlated, and the pure SU(2) rate $\Gamma_W$ is no longer strongly suppressed but reaches a plateau at small temperatures. Only the physically relevant combination $N_{\rm CS}^W - N_{\rm CS}^Y$ becomes frozen. The dashed vertical line at $T=145$\,GeV in Fig.~\ref{fig:su2_vs_su2u1} is the point where the measured rate matches the one from the pure SU(2) case. This is seen to happen systematically around $2$\,GeV below the pseudocritical temperature regardless of the magnitude of the magnetic flux $b$.

\begin{figure}[t]
  \includegraphics[width=1.0\columnwidth]{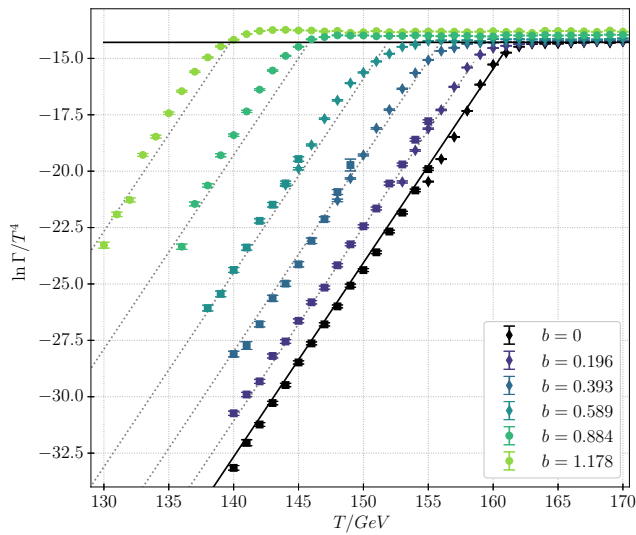}
  \caption{Sphaleron rate against temperature. Circle data points are from $V=32^3a^3$, diamond data points from $V=16^3a^3$ and square data points from multicanonical simulations. Grey dotted lines have the same slope as the fit of $b=0$ (black) and are shifted according to the shift of the pseudocritical temperature seen in Fig.~\ref{fig:Tcb}. Black horizontal line is the $b=0$ symmetric rate fit.}
  \label{fig:rate}
  \vspace{-0mm}
\end{figure}

The full diffusion rate \eqref{ncs_su2u1} is plotted against the temperature for different values of the magnetic field in Fig.~\ref{fig:rate}. (For clarity, we do not plot all values of $b$ that were simulated.) In the symmetric phase the SU(2) sphaleron rate is unaffected by the presence of the external magnetic field and the data is compatible with the $b=0$ case in Eq.~\eqref{sym_fit}.  However, the U(1) rate increases with increasing $b$, and so does the physically relevant $\Gamma$.  This is discussed in more detail below.

In the broken phase for small field values the slope at which the rate drops is compatible with the slope obtained from the $b=0$ fit. For larger magnetic field values we do not have enough data to verify this with confidence, but as shown in Fig.~\ref{fig:rate} the suppression of the rate continues to drop at approximately the same rate as at $b=0$, only the temperature is shifted to lower values.  The shift in temperature is roughly according to the shift in the pseudocritical temperature, as seen in Fig.~\ref{fig:Tcb}.  For the largest magnetic field we simulate, $b=1.178$ ($B_Y^{4d}/T^2 \approx 2.3$) the sphaleron rate suppression is shifted approximately to $22$\,GeV lower temperatures from the $b=0$ case.

\begin{figure}[t]
  \includegraphics[width=1.0\columnwidth]{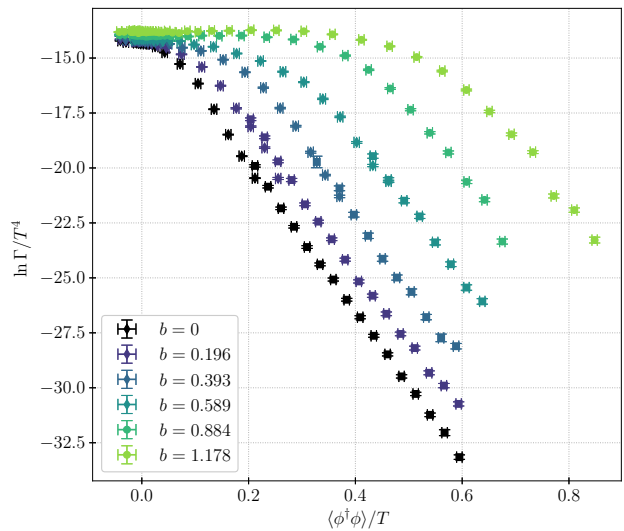}
  \caption{Sphaleron rate against the Higgs expectation value. The Higgs expectation value can get quite large before the rate gets suppressed.}
  \label{fig:rate_vev}
  \vspace{-0mm}
\end{figure}

The change in the sphaleron rate when the external field is increased can be understood to arise from two effects: the Higgs field expectation value decreases, and the sphaleron interacts with the field through its magnetic dipole moments.  Both of these effects reduce the sphaleron barrier.
To isolate the effects arising from the sphaleron dipole moment from the effects of changing Higgs expectation value we plot the sphaleron rate against the Higgs expectation value in Fig.~\ref{fig:rate_vev}.  It can be observed that at larger magnetic field values the Higgs expectation value can become quite large before the onset of the suppression of the rate.

\begin{figure*}
  \begin{centering}
    \includegraphics[width=1.0\columnwidth]{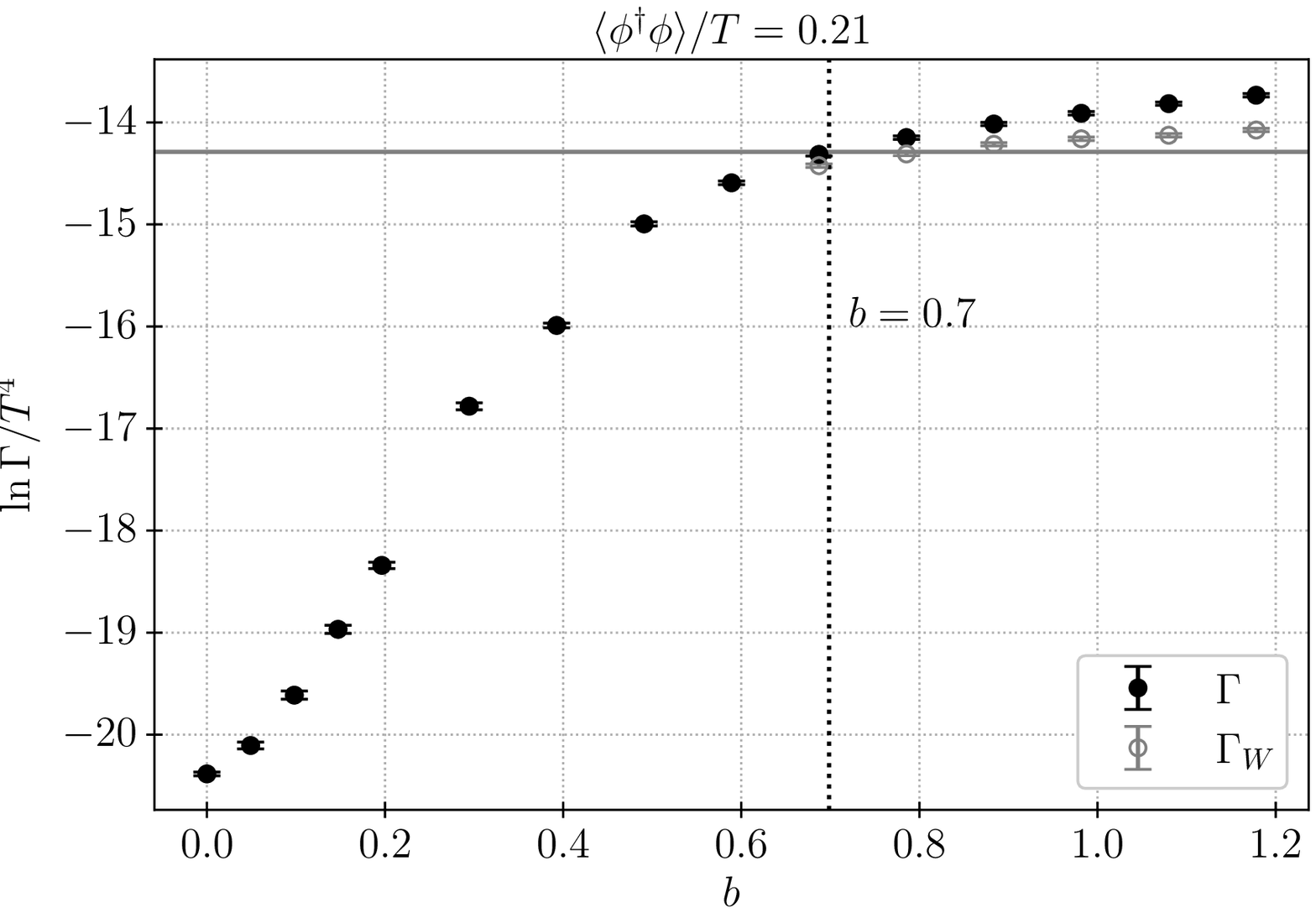}
    \includegraphics[width=1.0\columnwidth]{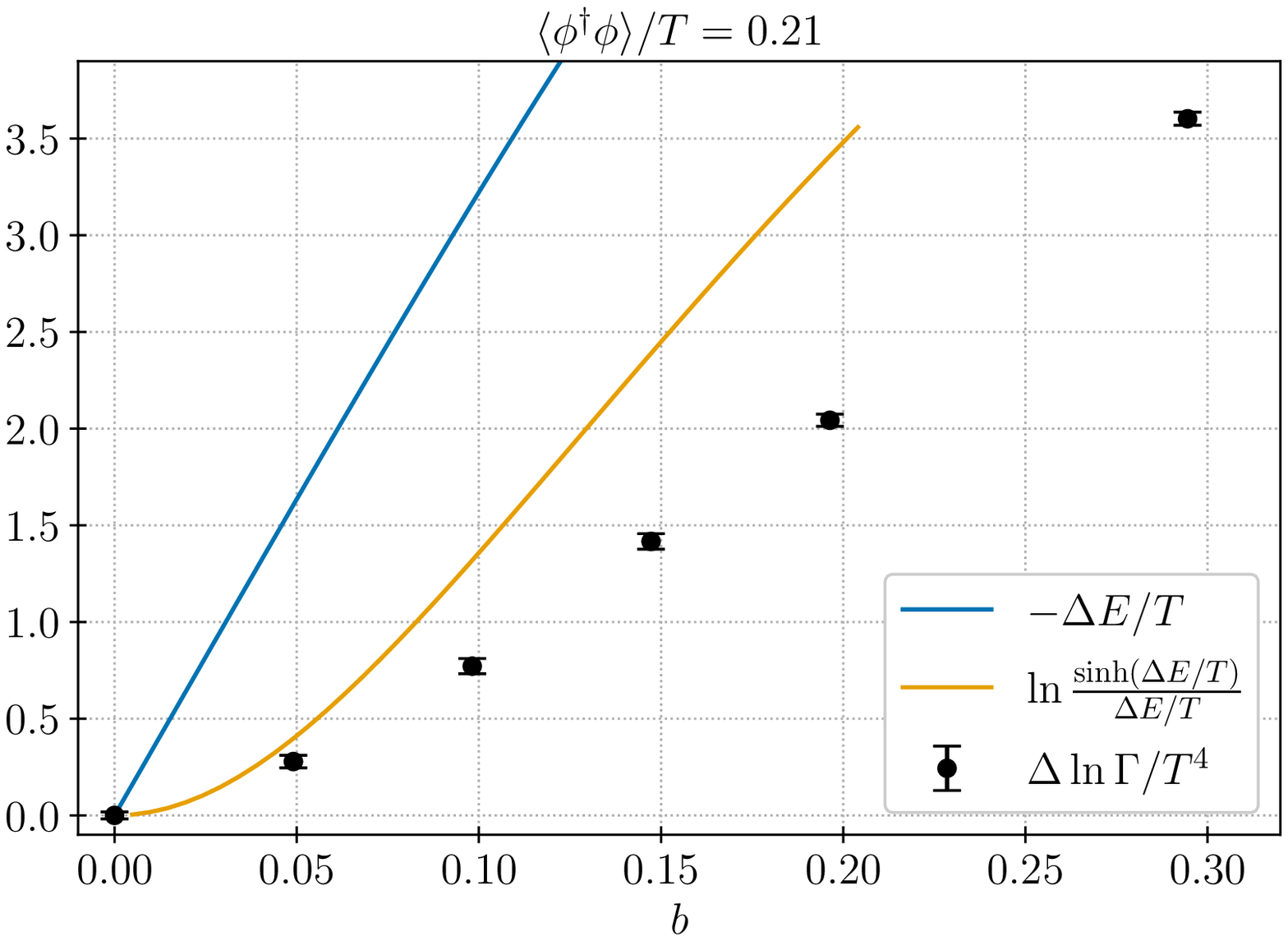}
  \end{centering}
  \caption{Left: the rate with constant Higgs expectation value. Vertical dotted line corresponds to the magnetic field value where the expectation value $\langle \phi^\dagger \phi \rangle/T=0.21$ is obtained at the pseudo critical temperature (i.e. to the right of the line we are getting in to the symmetric phase). The horizontal gray line is the symmetric $b=0$ sphaleron rate \eqref{sym_fit}. Right: comparing the difference of the rate $\Delta \ln \Gamma/T^4$ from simulations (black) to the analytical estimate (orange). We also plot the energy difference of the sphaleron configuration (blue) obtained from the analytical computation, see Appendix \ref{appendixA}.}
  \label{fig:const_phi_rate}
\end{figure*}

Finally, in Fig.~\ref{fig:const_phi_rate} we show how the sphaleron rate depends on the magnetic field at constant Higgs expectation value. This enables the comparison with the semianalytical results in Ref.~\cite{Comelli:1999gt}, where the change in the Higgs expectation value was neglected.
The rates at constant Higgs expectation value are obtained by interpolating the data shown in Fig.~\ref{fig:rate_vev}.  For small field values the rate behaves quadratically until around $b\simeq 0.2$ it reaches a linear regime. The linear regime ends when the field gets strong enough to start restoring the electroweak symmetry where the rate eventually reaches the $b=0$ SU(2) symmetric phase value \eqref{sym_fit}, see left plot in Fig.~\ref{fig:const_phi_rate}. 

\begin{figure*}[t]
  \begin{centering}
    \includegraphics[width=1.0\columnwidth]{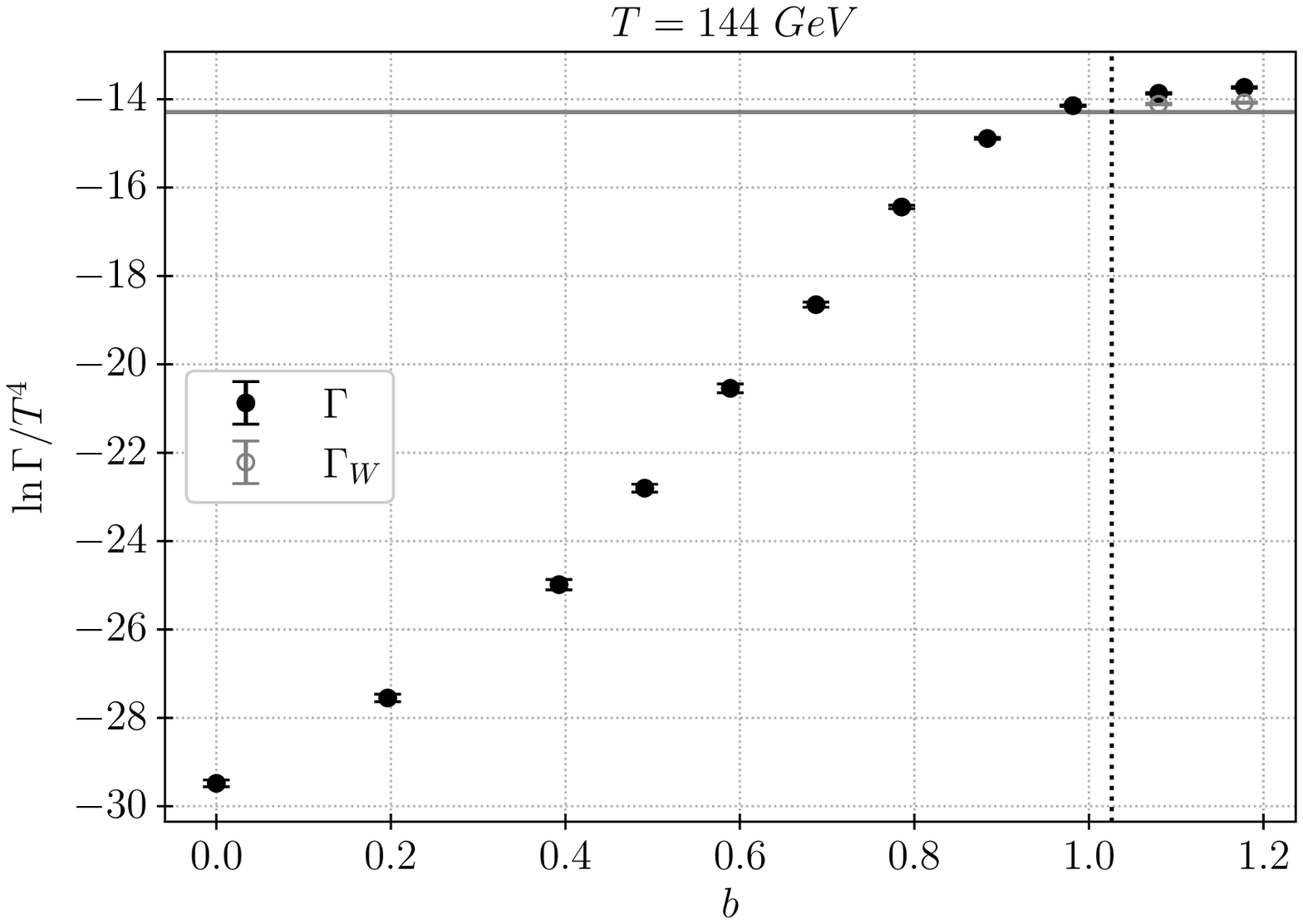}
    \includegraphics[width=1.0\columnwidth]{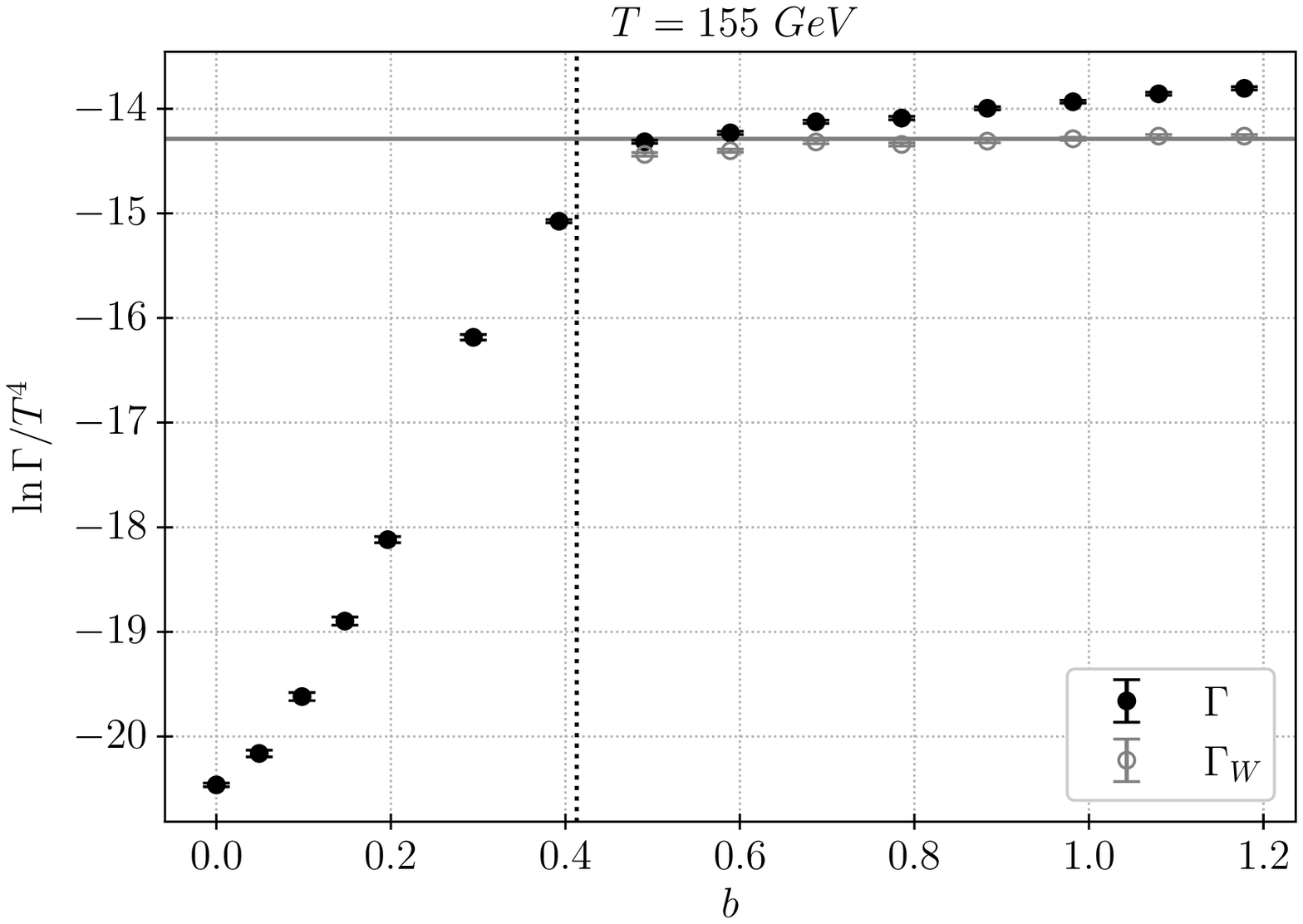}
  \end{centering}
  \caption{How the rate changes with the magnitude of the magnetic field with constant temperature $T=144$\,GeV on the left and $T=155$\,GeV on the right. The vertical dotted line is the value of $b$ where the constant temperature of the plot is the pseudocritical temperature. The horizontal gray line is the symmetric $b=0$ sphaleron rate \eqref{sym_fit}.}
  \label{fig:const_T_rate}
  \end{figure*}

Qualitatively similar behavior is seen when plotting the sphaleron rate with constant temperature, see Fig.~\ref{fig:const_T_rate}.  At small magnetic fields the change in $\ln \Gamma$  is proportional to $b^2$, turning into approximately linear behavior at intermediate $b$ until finally reaching the symmetric phase value where the rate flattens to constant.
Comparing Figs.~\ref{fig:phi2} and \ref{fig:rate_vev}, we can observe that the ``restoration'' of the rate happens before the Higgs field is fully restored.

To compare the simulation results to a semianalytical estimate we did the analysis presented in Ref.~\cite{Comelli:1999gt} (where they used nonphysical Higgs mass) but now with Standard Model parameters. Details of the computation can be found in Appendix \ref{appendixA}. From the analytical computation we get the sphaleron energy as a function of the external magnetic field. Assuming that for small fields the change in energy is due to a simple dipole interaction $\Delta E = -\vec{\mu}_{\text{sph}} \cdot \vec{B}^{4d}_c$ and that the change to the rate $\Delta \ln \Gamma/T^4 \equiv \ln \Gamma(b)/T^4-\ln \Gamma(b=0)/T^4$ is purely due to the change in energy, the change in the rate is approximately
\begin{equation}
    \Delta \ln \Gamma/T^4 \sim \ln\left[\frac{\sinh(\Delta E/T)}{\Delta E/T}\right].
    \label{eq:deltagamma}
\end{equation}
In Ref.~\cite{Comelli:1999gt} it was assumed that the change in $\ln\Gamma$ is directly proportional to the change of the minimum energy for the sphaleron, i.e. $\Delta \ln\Gamma/T^4 \propto \Delta E/T$. Our result in Eq.~\eqref{eq:deltagamma} takes into account the random orientations of the magnetic dipoles at finite $T$.  At small fields Eq.~\eqref{eq:deltagamma} gives $\Delta \ln \Gamma/T^4 \sim (\Delta E/T)^2/6$, and turns into $\sim$ linear behavior at larger field values.

For small external field values Eq.~\eqref{eq:deltagamma} is close to what we obtain from our simulations, however, the simple dipole approximation quickly becomes invalid, see Fig.~\ref{fig:const_phi_rate}.
Our results above indicate that at small magnetic fields the dominant effect on the sphaleron rate arises from the magnetic dipole moment of the sphaleron, and the change in the Higgs expectation value is subleading.

\begin{figure*}[t]
  \begin{centering}
    \includegraphics[width=1.0\columnwidth]{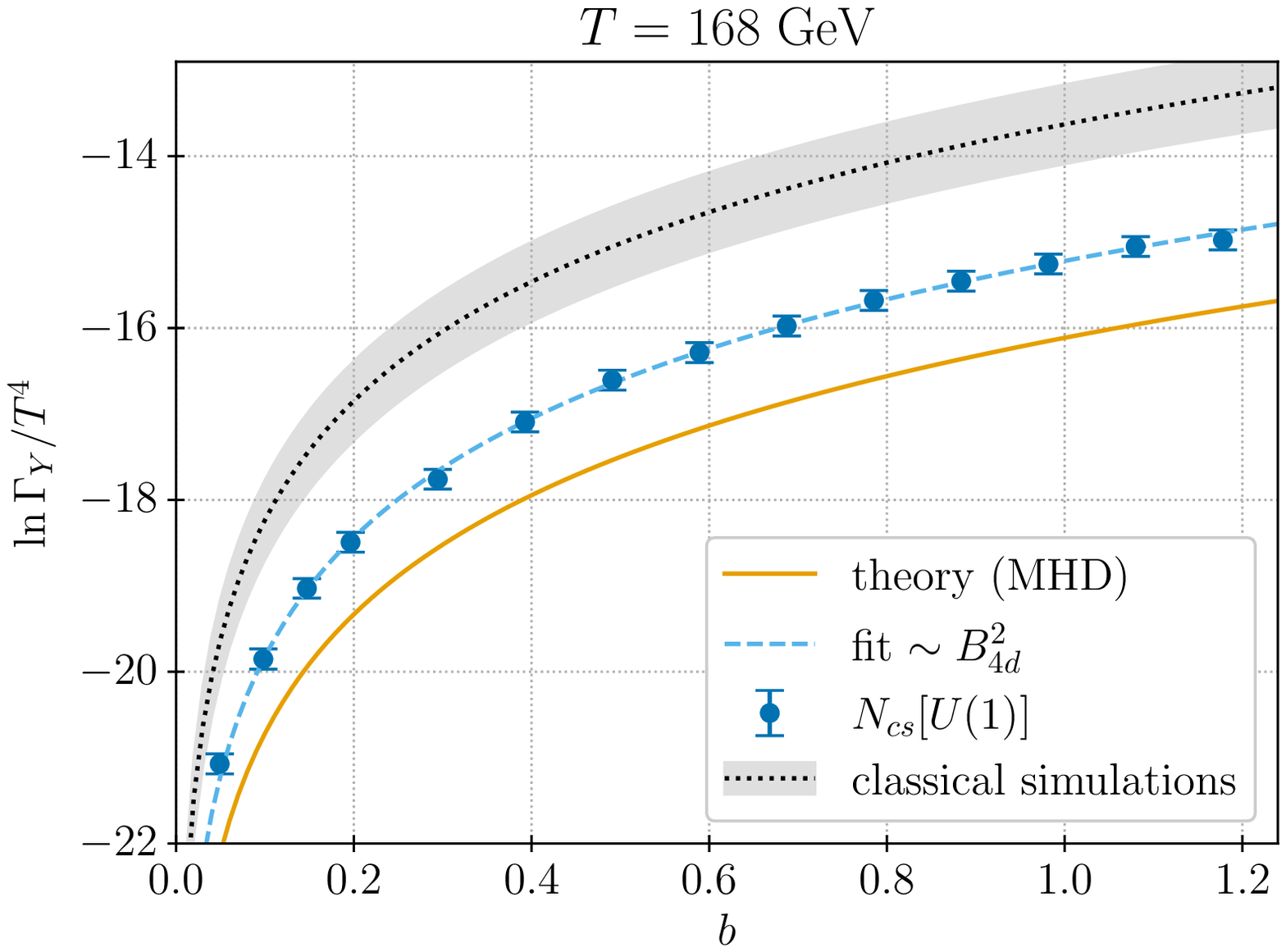}
    \includegraphics[width=1.0\columnwidth]{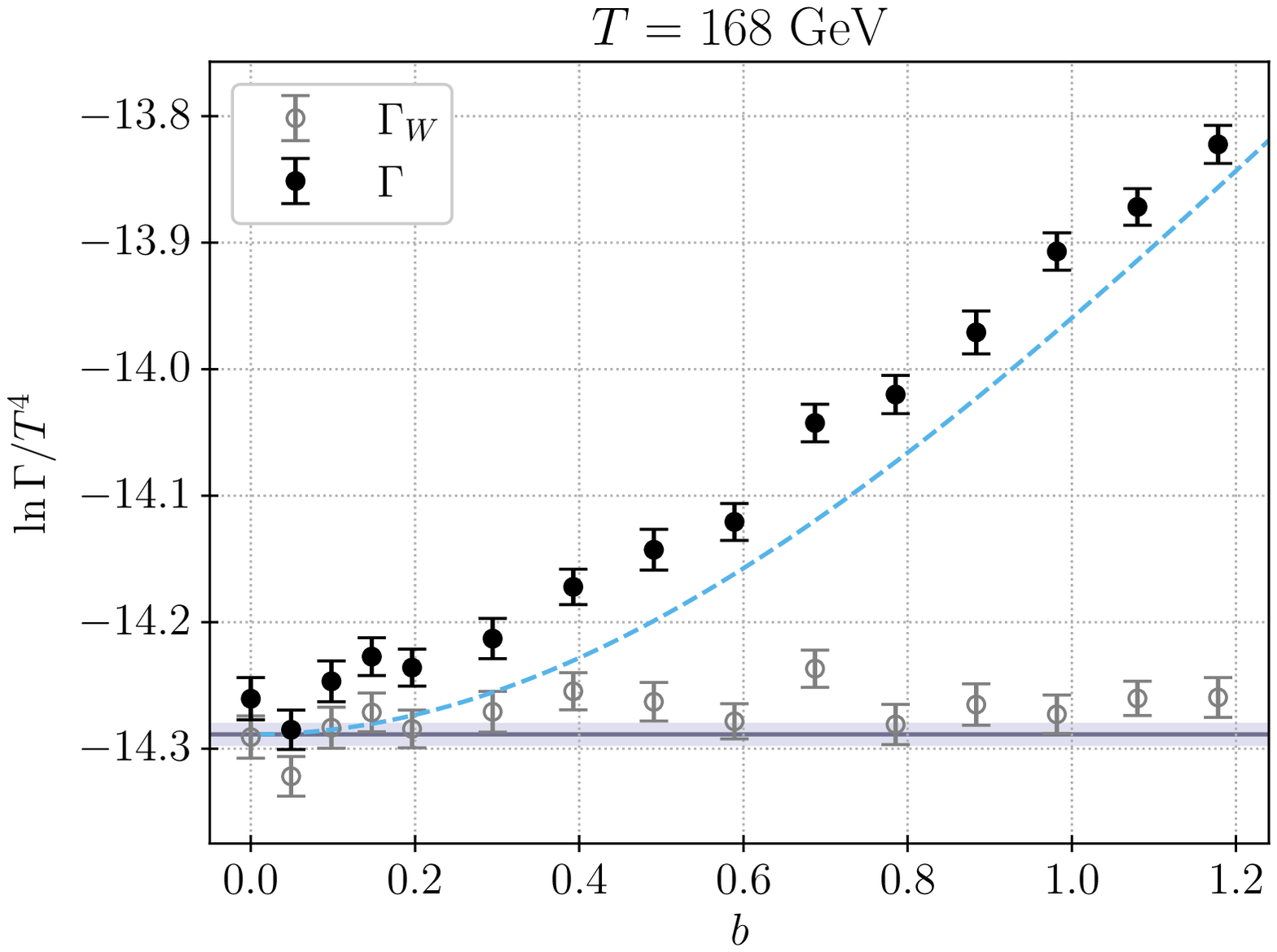}
  \end{centering}
  \caption{Left: diffusion rate of the U(1) CS number with different external magnetic fields (blue) compared with the results from classical simulations and expected rate from magnetohydrodynamics \cite{Figueroa:2017hun,Figueroa:2019jsi}. The shown rate is computed at $T=168$\,GeV, however, we do not find any systematic temperature dependence with the temperatures simulated. Right: pure SU(2) rate (gray) and the full rate (black) in the symmetric phase. The horizontal line is the $b=0$ SU(2) symmetric rate fit \eqref{sym_fit} and the blue dashed line is the sum of the latter and the fit from the left plot. The SU(2) rate $\Gamma_W/T^4$ stays approximately constant with increased magnetic field.}
  \label{fig:rate_u1}
\end{figure*}

Finally, let us look at the behavior of the sphaleron rate in the symmetric phase. Here the SU(2) rate does not show any systematic dependence on the magnetic field, see right plot in Fig.~\ref{fig:rate_u1}. Only the U(1) rate is affected by the magnetic field in the symmetric phase.  Despite the ambiguities associated with the U(1) field evolution in the symmetric phase as discussed in Sec. \ref{sec:evolution}, we investigate the U(1) rate in our simulations and the dependence on the magnetic field fits very well the expected $B_{4d}^2$ behavior \cite{Figueroa:2017hun}, see Fig.~\ref{fig:rate_u1}. As seen in Fig.~\ref{fig:su2_vs_su2u1} the U(1) rate $\Gamma_Y/T^4$ is approximately constant over temperature and we obtain a fit $\Gamma_Y/T^4 = (0.5 \pm 0.01)\times 10^{-3} g^\prime{}^6 B_{4d}^2$ (with $g^\prime{}^2\simeq0.12237$). Comparing this to results obtained from classical simulations of U(1)-Higgs theory (scalar QED) performed in \cite{Figueroa:2017hun,Figueroa:2019jsi} our rate is $\sim 4$ times slower; comparing with magnetohydrodynamics our rate is $\sim 3$ times faster.  Given the ambiguities in the update algorithm the qualitative agreement between the results is good.

\section{Conclusion}\label{sec:conc}

Using lattice simulations of an effective 3d theory of the Standard Model we have computed the baryon violation (sphaleron) rate over the electroweak crossover deep into the broken phase with an external magnetic field. Both the baryon violation rate and the form of the electroweak crossover is changed due to an external magnetic field.  We have argued that the fully dissipative Langevin-type update is accurate to leading logarithmic order in $g_W^2$ in the broken phase.

For zero external field we computed the rate with and without the U(1) fields included and found no difference between the results. The zero external field results differ slightly from previous results \cite{D'Onofrio:2014kta}, see \eqref{sym_fit} and \eqref{brk_fit} for our results. The difference is due to us using an updated value for the top mass (which affects the mapping between the physical and the effective 3d theory parameters) and the fact that the previous computations did not fully implement the partial $O(a)$ improvement.  Nevertheless, the difference is well within the uncertainties of the calculation.

The baryon violation rate is affected by an external magnetic field due to multiple factors. The sphaleron has a dipole moment and its energy can be lowered. With an external field also the U(1) contributes to the baryon violating rate. With an external field the combination of the SU(2) and U(1) CS numbers couple to the baryon violating current and in the broken phase it is precisely this combination that gets suppressed.

To get a picture of how the electroweak transition is affected by an external magnetic field we computed the Higgs expectation value and its susceptibility. As the magnetic field is increased, the crossover shifts to lower temperatures and the transition region broadens.  This shifts onset of the suppression of the sphaleron rate to lower temperatures. In the broken phase the rate increases with the external magnetic field. For small fields it increases quadratically before switching over to a linear regime. The linear regime stops after the field becomes strong enough to restore the electroweak symmetry where the rate reaches the symmetric phase value. 

For small external fields we performed a semianalytical computation for the sphaleron energy in an external field (following \cite{Comelli:1999gt}) and used a simple dipole approximation to estimate the change in the sphaleron rate. For small field values the semianalytical result and our simulations are in relatively good agreement, see Fig.~\ref{fig:const_phi_rate}. This shows that for small fields the sphaleron dipole moment has the biggest effect on the rate. However, for larger fields the simple dipole approximation quickly becomes invalid and nonlinear effects become important. 

In the symmetric phase the SU(2) and U(1) Chern-Simons numbers evolve independently. There are ambiguities in how to perform real-time lattice simulations of the U(1) field evolution in the symmetric phase. However, from our simulations we find no significant effect of the magnetic field on the pure SU(2) rate which behaves as $\propto T^4$ and is compatible with the zero external field value \eqref{sym_fit}. The U(1) part of the rate is found to increase with the magnetic field with the expected behavior $\propto B_{4d}^2$.  

Full results of our simulations are available as tables at Zenodo \cite{zenodo}.

\section*{Acknowledgments}

The authors acknowledge the support from the Academy of Finland Grants No. 345070 and No. 319066.
Part of the numerical work has been performed using the resources at the Finnish IT Center for Science, CSC.

\appendix 

\section{Continuum to lattice parameters with improvements}\label{appendix_cont_rel}

We use the partial $O(a)$ improvements computed in \cite{Moore:1996bf, Moore:1997np} (these are partial since there is an additive correction to the parameter $y$ that has not been computed to date). We choose the desired values of $x,y,z$ and $g_3^2 a$ that we want to simulate and then compute the relevant counterterms given by
\begin{align}
  Z_g^{-1} &= 1 + \frac{g_3^2 a}{4 \pi}\left(  \frac{\pi}{3} + 6\xi + \frac{\Sigma}{24}  \right), \label{def_Zg}\\
  Z_b &= 1 + z\frac{g_3^2 a}{4\pi} \left( \frac{\pi}{3} - \frac{\xi}{12} + \frac{\Sigma}{24} \right), \\
  Z_m^{-1} &=  1 + \frac{g_3^2 a}{4\pi}\left[ (9 - 24x + 3z )\frac{\xi}{4} + (3 + z)\frac{\Sigma}{24} \right],  \label{def_Zm}\\
  \delta x &= \frac{g_3^2 a}{4\pi}\Big\{ \Big[1 -6x(3+z)+48x^2 +\tfrac{1}{2}(1+z)^2\Big]\frac{\xi}{4} \nonumber\\
  &~~~~~~~~~~~ - x(3+z)\frac{\Sigma}{12} \Big\},
\end{align}
where $\Sigma = 3.175911...\ $ and $\xi=0.152859...\ $ are constants. We then construct the lattice action with the relations between the continuum parameters and the lattice parameters $\beta_G, \beta_Y, \beta_H, \beta_2, \beta_4$ \cite{Laine:1997dy}
\begin{align}
  \beta_Y &= \frac{\beta_G}{\bar{z}}\ , \beta_H  = \frac 8 {\beta_G}, ~ \beta_4  =  \frac{\beta_H^2}{\beta_G}\bar{x}, \label{latcont1} \\
  \frac{\beta_2}{\beta_H}  &=   3 + \frac{8\bar{y}}{\beta_G^2}
            - (3 + 12 \bar{x} + \bar{z})\frac{\Sigma}{4\pi\beta_G} \nonumber \\
            &-\frac{1}{2\pi^2\beta_G^2} \bigg[ 
             \left(
               \frac{51}{16} - \frac{9\bar{z}}8 - \frac{5\bar{z}^2}{16} 
               +9\bar{x} - 12\bar{x}^2 + 3\bar{x} \bar{z} \right)  \nonumber \\
            & ~~~~~~~\times\big(\ln(\tfrac{3}{2}\beta_G) + 0.09\big) \nonumber \\
           &+ 4.9 - 0.9\bar{z} + 0.01\bar{z}^2 + 5.2\bar{x} + 1.7\bar{x} \bar{z}\bigg],
            \label{latcont2}
\end{align}
using the modified parameters
\begin{align}
  \beta_G &= \frac{4}{g_3^2 a} Z_g^{-1} = \frac{4}{g_3^2 a} + 0.6674... \ , \\
  \bar{x} &= \frac{x + \delta x}{Z_g} ,\ \bar{y} = y\frac{Z_m}{Z_g^2} ,\ \bar{z} = z\frac{1}{Z_g Z_b} \ . \label{n_xyz}
\end{align}
Then the lattice observables are related to the continuum values with the parameters $x,y,z$ by a multiplicative correction (and possible renormalization factors). For example, the lattice observable $\langle \half\tr\Phi^\dagger\Phi \rangle$ is related to the $\msbar$ renormalized 3d continuum value $\langle\phi^\dagger\phi\rangle$ by Eq. \eqref{rl2}.

\section{Small field analytical estimate}\label{appendixA}

In this appendix we present details on the analytical computation in a small external field. We follow the analysis performed in \cite{Comelli:1999gt}. 

When the U(1) field is included the sphalerons spherical symmetry is reduced into an axial symmetry. With the physical value for the weak mixing angle $\theta_W$ the angular dependence of the solution is found to be mild \cite{Kunz:1992uh} at zero magnetic field. The expansion parameter, with external magnetic field $B^{4d}_c$, is effectively $\theta_W B^{4d}_c/gv^2$ and the angular dependence becomes relevant for larger magnetic fields. For small fields it suffices to use simpler ansatz that is spherically symmetric \cite{Klinkhamer:1990fi}. The ansatz depends on four functions $f(\xi),f_0(\xi),f_3(\xi),h(\xi)$ of dimensionless radial coordinate $\xi\equiv gvr$, where $v(T)$ is the temperature dependent Higgs expectation value. The energy functional of the sphaleron using the ansatz (see \cite{Klinkhamer:1990fi}\cite{Comelli:1999gt}) in a constant external hypermagnetic field $B^{4d}_c$ is $E = E_0 - E_{\text{dip}}$ with
\begin{align}
  E_0 &= \frac{4\pi v}{g}\int_0^\infty \rmd \xi \Bigg[ \frac{8}{3}f^\prime{}^2 + \frac{4}{3}f_3^\prime{}^2 + \frac{1}{2}\xi^2 h^\prime{}^2 + \frac{4g^2}{3g^\prime{}^2}f_0^\prime{}^2\nonumber \\
  &+ \frac{8}{3\xi^2}\left\{ 2f_3^2(1-f)^2 + [f(2-f)-f_3]^2 +\frac{g^2}{g^\prime{^2}}(1-f_0)^2\right\} \nonumber \\
  &+\frac{h^2}{3}\left\{ (f_0-f_3)^2 + 2(1-f)^2 \right\} +\frac{\lambda}{4g^2}\xi^2(h^2-1)^2 \Bigg] \ ,
\end{align}
and
\begin{align}
  E_{\text{dip}} &= \int_0^\infty \rmd \xi \frac{8\pi }{3g g^\prime v} \left[ -2\xi f_0^\prime + 2(1-f_0) \right] B^{4d}_c \ .
\end{align}
The field equations for the ansatz functions turn out as
\begin{align}\label{sph_odes}
  &h^{\prime\prime} + \frac{2}{\xi}h^\prime - \frac{2h}{3\xi^2}\left[ 2(1-f)^2 +(f_0-f_3)^2 \right] - \frac{\lambda}{g^2}(h^2-1)h = 0 , \nonumber\\
  &f^{\prime\prime} + \frac{1-f}{\xi^2}\left[2f(f-2)+2f_3+2f_3^2\right]+\frac{1}{4}(1-f)h^2=0 , \nonumber\\
  &f_0^{\prime\prime}+\frac{2(1-f_0)}{\xi^2}-\frac{g^\prime{}^2}{4g^2}(f_0-f_3)h^2 = 0, \nonumber\\
  &f_3^{\prime\prime}-\frac{2}{\xi^2}\left[ 3f_3+f(f-2)(1+2f_3) \right] - \frac{h^2}{4}(f_3-f_0) = 0 .
\end{align}
The ansatz functions are subject to the following boundary conditions
\begin{align}
  &f,h \to 1,\ f_3,f_0 \to 1- \frac{\sin 2\theta_W\xi^2}{8gv^2}B^{4d}_c,~~ \text{as}\  \xi \to \infty \  , \nonumber \\
  &f,h,f_3 \to 0, f_0 \to 1, ~~ \text{as}\ \xi \to 0 \ .
\end{align}
It is convenient to make a change of variables 
\begin{equation}
  g_i(\xi) = f_i(\xi) + \sin 2\theta_W \frac{\xi^2}{2gv^2},
\end{equation}
for $i = 0,3$, so that the boundary conditions for the new functions at infinity are simply $g_i \to 1$ as $\xi\to\infty$. Furthermore, we use a change of variables $x \equiv \xi/(3+\xi)$ which maps $\xi\to\infty$ to $x\to 1$. Finally we use the standard model values for the parameters $\lambda,g,g^\prime$ at the electroweak scale.

\begin{figure}[t]
  \includegraphics[width=1.0\columnwidth]{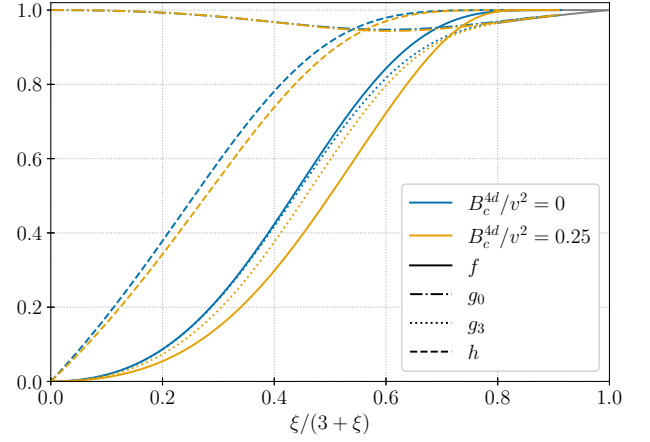}
  \caption{Examples of numerical solutions for the functions $f,h,g_0,g_3$ for zero and non-zero magnetic field.}
  \label{fig:spherical_sph_func}
  \vspace{-0mm}
\end{figure}

With the above we have all the ingredients to compute the change to the sphaleron energy for small fields in the spherical approximation. From the set of coupled differential equations \eqref{sph_odes} we solve the functions $f,h,g_0,g_3$ numerically using a fourth order collocation method implemented in \cite{2020SciPy-NMeth} . 
Equations \eqref{sph_odes} are divergent at the boundaries and thus we solve the system only in range $[\epsilon_1,1-\epsilon_2]$ where the $\epsilon_i$ are small offsets. Despite the divergences in the equations the solutions are completely regular near the boundaries and we just linearly extrapolate them to boundary values. The accuracy of this simple procedure is  sufficient for our comparison purposes. A set of solved functions is plotted in Fig.~\ref{fig:spherical_sph_func} for zero magnetic field and for one example of a nonzero magnetic field.  Typical range of $[\epsilon_1, 1-\epsilon_2]$ is $[0.0006,0.913]$, but even large variations of these values does not significantly change the solutions or the energy computed from them. 

\begin{figure}[t]
  \includegraphics[width=1.0\columnwidth]{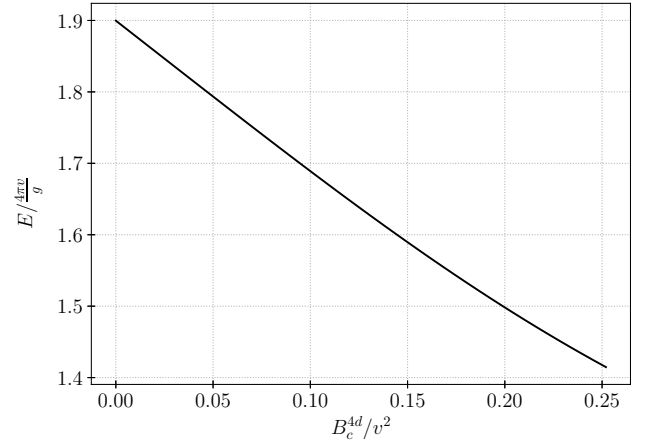}
  \caption{Energy of the sphaleron configuration (computed using the spherical approximation) in terms of the magnitude of the magnetic field.}
  \label{fig:spherical_sph_E}
  \vspace{-0mm}
\end{figure}

The energy of the sphaleron configuration is obtained by numerically integrating over the energy functional while omitting the constant external magnetic field terms which would make the expression divergent. The energy as a function of the magnetic field is plotted in Fig.~\ref{fig:spherical_sph_E}.
Assuming that the change in energy $\Delta E \equiv E(B=0)-E(B)$ is due to a simple dipole interaction $\Delta E = -\vec{\mu}_{\text{sph}} \cdot \vec{B}^{4d}_c$ and that the change of the rate $\Gamma$ is only due to this energy difference $\Gamma \sim \exp(\Delta E / T) \Gamma_0$, where $\Gamma_0$ is the rate without magnetic field. Averaging over the space of orientations for the dipole the change to the rate is roughly
\begin{align}
  \Delta\ln\Gamma/T^4 &\sim \ln\left\{\int \frac{\rmd\Omega}{4\pi} \exp\left[-\frac{\mu_{\text{sph}}B^{4d}_c}{T} \cos\theta\right]\right\} \nonumber \\ 
  &\simeq \ln\frac{\sinh(\Delta E / T)}{\Delta E / T} \ .
\end{align}

\bibliography{main}

\end{document}